\documentclass[preprint,prb]{revtex4}

\usepackage{amsmath}

\usepackage{amssymb}
\usepackage{amsthm}

\newcommand{{\bff}}{\mbox{\boldmath$f$\unboldmath}}

\newcommand{{\bfF}}{\mbox{\boldmath$F$\unboldmath}}

\newcommand{{\bfA}}{\mbox{\boldmath$A$\unboldmath}}

\newcommand{\gradv}{{\nabla}}
\def\v#1{{\bf#1}}
\begin{document}

\title{\Large A short proof that the Coulomb-gauge \\potentials yield the retarded fields}

\author{Jos\'e A. Heras}
\email{herasgomez@gmail.com}
\affiliation{Universidad Aut\'onoma Metropolitana, Unidad Azcapotzalco, Av. San Pablo No. 180, Col. Reynosa, 02200, M\'exico D. F. M\'exico}
\begin{abstract}
A short demonstration that the potentials in the Coulomb gauge yield the retarded electric and magnetic fields is presented. 
This demonstration is relatively simple and can be presented in an advanced undergraduate curse of electromagnetic theory,
\end{abstract}
\maketitle
\noindent{\bf 1. Introduction}
\vskip 10pt
\noindent
After the Lorenz gauge, the most popular gauge in electrodynamics is the Coulomb gauge in which the scalar potential $\Phi_C$ displays the properties of acausality and instantaneous propagation. However, the electric field generated by the Coulomb-gauge potentials $\Phi_C$ and $\v A_C$ is a retarded field satisfying the properties of causality and propagation at speed of light $c$. The proof of this well-known result seems to be, however, rather involved. Most textbooks usually do not present this proof and in the best of cases some of them quote a paper of Brill and Goodman [1] who presented in the 1960's an elaborate demonstration (restricted to sources with harmonic time dependence) that potentials in the Coulomb and Lorenz gauges yield the same retarded fields. Starting the 2000's, Jackson [2] derived a novel expression for $\v A_C$ and demonstrated how this 
expression and the usual expression for the potential $\Phi_C$ lead to the retarded fields. However, Jackson's formula for $\v A_C$ is not so easy to obtain. More recently, other authors [3-7] developed alternative procedures to show that the potentials $\Phi_C$ and $\v A_C$ generate the well-known retarded fields. 

In this note I present a short proof that the Coulomb-gauge potentials yield the retarded electric and magnetic fields. The quantities $-\gradv\Phi_C$ and $-\partial \v A_C/\partial t$ are expressed in terms of retarded integrals with local and non-local sources. The combination  $-\gradv\Phi_C-\partial \v A_C/\partial t$ identically cancels the integrals with non-local sources and identifies with the retarded electric field in SI units. The retarded magnetic field is directly obtained by taking the curl to $\v A_C$. This short proof involves simple vector operations on retarded quantities and may be presented in an advanced undergraduate curse of electromagnetic theory. To emphasize the simplicity of this proof, it is compared with the more elaborated proof given by Jackson [2]. 

\vskip 12pt
\noindent{\bf 2. The short proof}
\vskip 10pt
\noindent
The electric and magnetic fields expressed in terms of the Coulomb-gauge potentials  are
\begin{align}
\v E=&-\gradv\Phi_C-\frac{\partial \v A_C}{\partial t},\\
\v B=&\gradv\times\v A_C.
\end{align}
The equations for the potentials $\Phi_C$ and $\v A_C$ are given by
\begin{align}
\nabla^2\Phi_C=&-\rho/\epsilon_0,\\
\Box^2\v A_C=&-\mu_0\v J+ \frac{1}{c^2}\frac{\partial }{\partial t}\gradv\Phi_C,
\end{align}
where $\Box^2 \equiv \nabla^2 - (1/c^2)\partial^2/\partial t^2$. The retarded solution of (4)  can be written as$^1$\footnotetext[1]{To verify that (5) satisfies (4) one takes $\Box^2$ to (5), uses the general identity $\Box^2[\;\;]/R=-4\pi[\;\;]\delta\{\v x-\v x'\}$ with $\delta$ being the Dirac delta function, and integrates over all space. The general identity is proved in 
Heras J A 2007 Can Maxwell's equations be obtained from the continuity equation?{\it Am. J. Phys.} {\bf 75} 652}
\begin{align}
\v A_C= \frac{\mu_0}{4\pi}\int d^3x'\frac{[\v J]}{R} -\frac{1}{4\pi c^2}\!\int \!d^3x'\frac{1}{R}\bigg[\frac{
\partial\gradv'\Phi_C}{\partial t'}\bigg],
\end{align}
where $R=|\v x-\v x'|$, the quantities in  the brackets $[\;]$ are evaluated at the retarded time $t'=t-R/c$ and the integral is over all space. 
From (5) and $\partial [\;\;]/\partial t=[\partial/\partial t']$ one obtains 
\begin{align}
-\frac{\partial \v A_C}{\partial t}= -\frac{\mu_0}{4\pi}\int d^3x'\frac{[\partial\v J/\partial t']}{R} +\frac{1}{4\pi c^2}\!\int \!d^3x'\frac{1}{R}\bigg[\frac{
\partial^2\gradv'\Phi_C}{\partial t'^2}\bigg].
\end{align}
 Consider now the identity
\begin{align}
-\Box^2\gradv \Phi_C\!\equiv -\nabla^2\gradv \Phi_C
+\frac{1}{c^2}\frac{\partial^2}{\partial t^2}\gradv \Phi_C.
\end{align}
Equations (3) and (7) and the result $\nabla^2\gradv  \Phi_C=\gradv\nabla^2 \Phi_C$ yield the following wave equation 
\begin{align}
-\Box^2\gradv\Phi_C=& \gradv\rho/\epsilon_0+\frac{1}{c^2}\frac{\partial^2}{\partial t^2}\gradv\Phi_C.
\end{align}
The retarded solution of this equation is given by$^{2}$ \footnotetext[2]{Strictly, (9) is not a solution for $-\gradv\Phi_C$, but only an integral representation given in terms of its second-order time derivatives and of the charge density. 
One can verify that (9) satisfies (8) by taking $\Box^2$ to (9), using $\Box^2[\;\;]/R=-4\pi[\;\;]\delta\{\v x-\v x'\}$ and integrating over all space}
\begin{align}
-\gradv\Phi_C= -\frac{1}{4\pi\epsilon_0}\int\!d^3x'\frac{[\gradv'\rho]}{R}-\frac{1}{4\pi c^2}\!\int \!d^3x'\frac{1}{R}\bigg[\frac{
\partial^2\gradv'\Phi_C}{\partial t'^2}\bigg].
\end{align}
Equations~(6) and (9) show that each of the terms $-\partial \v A_C/\partial t$ and  $-\gradv\Phi_C$ can be expressed as retarded integrals with local and non-local sources. The second integral in (9) exactly cancels the second integral in (6). Therefore, if (6) and (9) are used in (1) then one obtains the electric field expressed in its usual retarded form [8]:
\begin{align}
\v E=-\frac{1}{4\pi\epsilon_0}\int d^3x'\frac{[\gradv'\rho+(1/c^2)\partial\v J/\partial t']}{R}.
\end{align}
Curl of (5) and an integration by parts give the magnetic field in its usual retarded form:
\begin{align}
\v B= \frac{\mu_0}{4\pi}\int\!d^3x'\gradv \times\frac{[\v J]}{R}=\frac{\mu_0}{4\pi}\int\!d^3x'\frac{[\gradv'\times \v J]}{R}.
\end{align}
The above proof that $\Phi_C$ and $\v A_C$ yield the retarded fields is relatively simple and could be presented in an advanced undergraduate curse of electromagnetism.
\vskip 10pt
\noindent{\bf 3. Jackson's proof}
\vskip 10pt
\noindent
Jackson's proof that the Coulomb-gauge potentials yield the retarded electric field can be outlined as follows. The instantaneous solution of (3) reads
\begin{align}
\Phi_C(\v x,t)=\frac{1}{4\pi\epsilon_0}\int\!d^3x'\frac{\rho(\v x',t)}{R}.
\end{align}
The gradient of (12) and an integration by parts allow one to express the field $\v E$ in (1) as
\begin{align}
\v E=\frac{1}{4\pi\epsilon_0}\int d^3x'\frac{\rho(\v x',t)\hat{\v R}}{R^2} -\frac{\partial \v A_C}{\partial t}.
\end{align}
The instantaneous (first) term in (13) must be a spurious quantity because the field $\v E$ is expected to be a retarded field.$^{3}$ \footnotetext[3]{
Maxwell's equations for sources confined in space and bounded in time admit retarded and advanced solutions (or more in general a combination of them). One chooses the retarded solutions because they satisfy the causality principle. See, Rohrlich F 2002 Causality, the Coulomb field, and Newton's law of gravitation {\it Am. J. Phys.} {\bf 70} 411-414} This means that the term $-\partial\v A_C/\partial t$ must contain a piece that identically eliminates 
the first term in (13). Jackson [2] has constructed an expression for $\v A_C$ using an indirect approach based on the gauge function that transforms the Lorenz-gauge potentials into the Coulomb-gauge potentials [2]:
\begin{equation}
\v A_C(\v x,t)= \frac{\mu_0}{4\pi}\!\int\!d^3 x'\frac{1}{R}\Big([\v J-c\hat{\v R}\rho] 
+ \frac{c^2\hat{\v R}}{R^2}\!\int_{0}^{R/c}\!d\tau\rho(\v x',t-\tau)\Big),
\end{equation}
where $\hat{\v R}\!=\!{\v R}/R\! =\!(\v x\!-\!\v x')/|\v x\!-\!\v x'|$. Time derivative of (14) and the property $\partial/\partial t=-\partial/\partial\tau$ yield the expected result [2]:
\begin{align}
-\frac{\partial \v A_C}{\partial t}=\frac{1}{4\pi\epsilon_0}\int d^3x'\bigg(\frac{\hat{\v R}[\rho]}{R^2}\!+\!\frac{\hat{\v R}[\partial \rho/\partial t']}{Rc}
\!-\!\frac{[\partial \v J/\partial t']}{Rc^2}\bigg)
-\frac{1}{4\pi\epsilon_0}\int d^3x'\frac{\rho(\v x',t)\hat{\v R}}{R^2}.
\end{align}
The last term in (15) exactly cancels the first instantaneous piece in (13). Therefore, from (13) and (15) one obtains 
the retarded electric field in the form given by Jefimenko [8]:
\begin{equation}
\v E=\frac{1}{4\pi\epsilon_0}\int d^3x'\bigg(\frac{\hat{\v R}[\rho]}{R^2}\!+\!\frac{\hat{\v R}[\partial \rho/\partial t']}{Rc}
\!-\!\frac{[\partial \v J/\partial t']}{Rc^2}\bigg).
\end{equation}
The curl of (14) gives the magnetic field expressed in its usual retarded form [8] (this calculation is similar to that in (11) because of the result $\gradv\times \hat{\v R}=0$). The practical difficulty in Jackson's proof is that the derivation of the expression for $\v A_C$ in (14) is somewhat laborious [2].

\vskip 20pt
\noindent{\bf 4. Comparing the two proofs}
\vskip 10pt
\noindent
The short proof that $\Phi_C$ and $ \v A_C$ yield the retarded electric field can be drawn as follows:
\begin{align}
    &\qquad\qquad\qquad\quad\;-\gradv\Phi_C\qquad\qquad\qquad\qquad\qquad\qquad\qquad\quad-\frac{\partial \v A_C}{\partial t}\nonumber\\
\v E=&\overbrace{\!-\frac{1}{4\pi\epsilon_0}\!\!\int\!\!d^3x'\frac{[\gradv'\rho]}{R}\!-\!\frac{\mu_0\epsilon_0}{4\pi}\!\!\int \!\!d^3x'\!\frac{1}{R}\!\bigg[\!\frac{
\partial^2\gradv'\Phi_C}{\partial t'^2}\!\bigg]
}\; \overbrace{ \!-\frac{\mu_0}{4\pi}\!\int\!\! d^3x'\frac{[\partial\v J/\partial t']}{R}\! +\!\frac{\mu_0\epsilon_0}{4\pi}\!\!\int\!\! \!d^3x'\frac{1}{R}\!\bigg[\!\frac{
\partial^2\gradv'\Phi_C}{\partial t'^2}\!\bigg]}.
\end{align}
The second and last terms identically cancel and therefore they represent spurious contributions which do not appear in the final expression for the retarded electric field. Jackson's proof that $\Phi_C$ and $ \v A_C$ yield the retarded electric field can be drawn as follows:
\begin{align}
   &\qquad\quad-\gradv\Phi_C\qquad\qquad\qquad\qquad\qquad\qquad\qquad\quad-\frac{\partial \v A_C}{\partial t}\nonumber\\
\v E=&\overbrace{\!\frac{1}{4\pi\epsilon_0}\!\int\! d^3x'\frac{\rho(\v x',t)\hat{\v R}}{R^2}}\;\overbrace{+\frac{1}{4\pi\epsilon_0}\!\int\!\! d^3x'\!\bigg(\!\frac{\hat{\v R}[\rho]}{R^2}\!+\!\frac{\hat{\v R}[\partial \rho/\partial t']}{Rc}
\!-\!\frac{[\partial \v J/\partial t']}{Rc^2}\!\bigg)\!-\!\frac{1}{4\pi\epsilon_0}\!\int\!\! d^3x'\frac{\rho(\v x',t)\hat{\v R}}{R^2}\!}.
\end{align}
The first and last terms identically cancel and therefore they represent spurious contributions which do not appear in the final expression for the retarded electric field. 

\vskip 10pt

{\it Note added in proof.} Note that (5), (6) and (9) are not solutions, but non-local coupled integral equations. To show
that the Coulomb-gauge potentials yield retarded fields one can equivalently use solutions (as in Jackson's approach)
or coupled integral equations (as in the present paper). Note also that the 'integral equation' in (5) can be transformed
into the 'solution' in (14) after some calculation. I thank Professor J D Jackson for calling my attention to this point.

\end{document}